\documentclass[useAMS,usenatbib]{mn2e}
\usepackage{txfonts}
\usepackage{natbib}
\usepackage{longtable}
\usepackage[dvips]{graphicx}
\usepackage{aas_macros}

\title[On the association of ULXs with young superclusters]{On the association of ULXs with young superclusters: M82 X-1 and a new candidate in NGC 7479}
\author[R. Voss et al.]{R. Voss$^{1}$, M. T. B. Nielsen$^{1}$, G. Nelemans$^{1}$,
M. Fraser$^{2}$, S. J. Smartt$^{2}$\\
$^{1}$ Department of Astrophysics/IMAPP, Radboud University Nijmegen, PO Box 9010, NL-6500 GL Nijmegen, the Netherlands\\
$^{2}$ Astrophysics Research Centre, School of Mathematics and Physics, Queen's
University Belfast, Belfast BT7 1NN, United Kingdom
}
\begin{document}

\date{}

\pagerange{\pageref{firstpage}--\pageref{lastpage}} \pubyear{2011}

\maketitle

\label{firstpage}

\begin{abstract}
We investigate the spatial coincidence of ultra-luminous X-ray sources
(ULXs) with young massive
stellar clusters. In particular we perform astrometry on
{\it Chandra} and {\it HST} data of two ULXs that are possibly
associated with such clusters. 

To date M82 X-1 is the only ULX claimed to be coincident with a
young massive stellar cluster. We remeasure the position of this source with
a high accuracy and find that the position of the X-ray source is 0.65
arcsec away from the stellar cluster, corresponding to an offset significance
of 3 sigma.  

We also report the discovery of a new candidate, based on observations
of NGC 7479. One of the ULXs observed in 
three X-ray observations is found
to be spatially coincident (within 1 sigma of the position error) 
with a young super-cluster observed in the
{\it HST} images. In the brightest state, the absorbed luminosity of the 
ULX is a few times $10^{40}$ erg s$^{-1}$, and in the faintest state below 
the detection limit of $\sim4$ times $10^{39}$ erg s$^{-1}$. The luminosity
in the brightest state requires an accreting black hole mass of at
least 100 M$_{\odot}$ assuming isotropic emission. However it is possible
that the source is contaminated by X-ray emission from the nearby supernova
SN2009jf. In this case the luminosity of the ULX is in a range where it
is strongly debated whether it is a super-Eddington stellar mass black
hole or an intermediate mass black hole.
The colours of the host cluster indicate a young stellar population, 
with an age between 10 and 100 Myr. The total stellar mass of the cluster 
is $\sim5\cdot10^{5}$M$_{\odot}$.

\end{abstract}

\section{introduction}
Ultra-luminous X-ray sources (ULXs) are defined as bright
($>10^{39}$ erg s$^{-1}$) X-ray sources located off-centre
in their host galaxies. Their luminosities are
above the Eddington limit for a $\sim$10M$_{\odot}$ black hole (BH)
assuming spherical accretion of hydrogen, approximately the highest
mass expected from normal stellar evolution in a solar
metallicity environment \citep[e.g.][]{Heger2003}. 
It is possible that such luminosities are caused by beaming
\citep{King2001}, or that systems achieve super-Eddington
accretion \citep{Begelman2002,King2003}. Alternatively, it means that 
the accreting objects are intermediate mass black holes (IMBHs) or 
supermassive black holes (SMBHs)
formed through other processes than stellar evolution.
The formation of such massive black holes is expected to take
place in massive stellar clusters, where dynamical effects cause
stars to sink to the centre \citep{Ebisuzaki2001,Portegies2002}. 
Alternatively, they can be formed in the central parts of the galaxy 
and be ejected by BH-BH interactions \citep[recoiling SMBHs][]{Bonning2007}. 
In recent years a sample
of extremely bright hyper-luminous X-ray sources (HLXs) have been
observed with X-ray luminosities in excess of 10$^{41}$ erg s$^{-1}$,
the brightest of which is $>10^{42}$ erg s$^{-1}$ \citep{Farrell2009},
strengthening the IMBH/SMBH interpretation, as such high apparent
luminosities are very difficult to achieve through anisotropic
radiation. 

The birth environment of ULXs can provide important
clues to their nature. Stellar mass black holes will be formed in any
region with star formation, although their formation could be diminished
in dense clusters due to the merger of their progenitors.
On the other hand, intermediate mass black holes are believed
to be formed from mergers of massive stars. They can therefore only
be formed in the dense environment of young massive star clusters.
It is therefore important to investigate the coincidence of ULXs
with stellar superclusters.
These clusters are the most massive young stellar clusters, with
masses in excess of a few times $10^{5}$M$_{\odot}$, and a very high 
specific stellar density. They are relatively rare and confined 
to galaxies with high star-formation rates 
(the Milky Way does not host any superclusters).
Optical observations of a large number of ULXs have yielded only three
possible associations. M82 X-1 is a ULX in M82 that has been associated with 
a cluster of mass of a few times 10$^{5}$M$_{\odot}$ and age of $\sim10$ Myr 
\citep{Portegies2004}. However, the astrometric accuracy of this study
was low, and below we re-analyze the association and show that there
is a significant offset between the X-ray source and the proposed host
cluster. The other two ULXs are located in the Antennae galaxies
\citep{Zezas2006}, but both were recently found to be located at
distances of $\sim0.6$ arcsec from the stellar clusters \citep{Clark2011},
corresponding to an offset significance of 2-3 sigma. This does not mean that the three ULXs are
not related to the stellar clusters, but rather that they were most
likely ejected from these relatively recently.

Other luminous sources have been claimed to be compatible with the position 
of a young cluster, but none have a color consistent with stellar ages 
below $10^8$ years \citep[e.g.][]{Ptak2006}. \\
In general ULXs are predominantly found in regions with high 
star-formation rates \citep{Irwin2004} and low metallicities
\citep{Pakull2002,Zampieri2009,Mapelli2010}. Observations of individual 
ULXs have found associations also with relatively poor 
stellar clusters \citep{Grise2008}, and in some cases with large
super-bubbles in the ISM \citep{Pakull2003}. This clearly shows that
ULXs are related to young stars, with ages of tens to a few hundred Myr.\\
However, a number of ULXs have been found to reside in globular 
clusters \citep{Maccarone2007,Maccarone2011}. These are likely BHs
created during the early evolution of the globular clusters.

\section{Revisiting M82 X-1}
The only other ULX that has been found to be spatially coincident
with a massive young cluster is M82 X-1, which is claimed to be
associated with the stellar cluster MGG-11. This was based on an analysis
with an astrometric accuracy of $\sim1$ arcsec \citep{Portegies2004}. 
Motivated by the results presented above, we revisit the astrometry of
this source. 
\begin{figure}
\resizebox{\hsize}{!}{\includegraphics[angle=0]{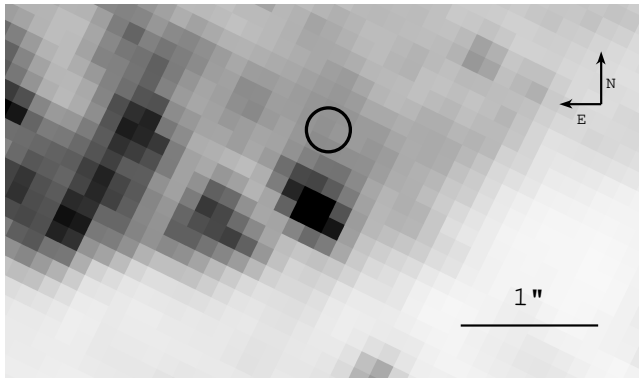}}
\caption{The position of M82 X-1 (circle indicating the 1 sigma
confidence region) compared to the young cluster MGG-11 (source below 
the circle). The position of the X-ray source is $\sim3$ sigma off-set
from the position of the cluster.}
\label{fig:M82}
\end{figure}
The region is covered well both by {\it Chandra} and {\it HST}. We select the
two deepest archival {\it Chandra} {\it ACIS} observations of M82 X-1 
(OBSID 10542 and OBSID 10543). With
$\sim120$ ks of exposure in each, they are deep enough to obtain
a precise astrometric solution. We have compared them to the
existing {\it HST} images in the vicinity of M82 X-1, and we find no
secure matches that can be used for a direct registration of the
images. Instead we register the X-ray images to {\it Sloan Digital
Sky Survey data release 8}\footnote{http://www.sdss3.org/} (SDSS), with 6
(5) matching sources in OBSID 10542 (10543). The results from the two
observations agree within 0.05 arcsec, and we use the
position from OBSID 10542 in the following.
We choose two {\it HST} observation sets that both have clear SDSS matches 
and where MGG-11 is clearly observed. Archival NICMOS F160W images obtained
in April 1998 were used for the original matching \citep{McCrady2003,Portegies2004}, and three matches 
are found in SDSS. As an alternative we used archival WFC3 F110W images 
obtained November 2009. For these we find four SDSS matches (of which only
two are the same as in the NICMOS images). Again the positions agree
within 0.05 arcsec. Based on these, we translate
the position of M82 X-1 into the {\it HST} images with a precision $\lesssim0.2$
arcsec. The results are shown in figure \ref{fig:M82}. We find that 
MGG-11 is located $\sim0.65$ arcsec south of M82 X-1 and that the
positions are inconsistent at the 3 sigma level.

\section{A new candidate in NGC 7479.}
NGC 7479 is located at a distance of ~33 Mpc 
(We adopt a distance modulus of 32.65, assuming a radial velocity 
corrected for infall onto Virgo of 2443 km s$^{-1}$ and a Hubble 
constant of 72 km s$^{-1}$ Mpc$^{-1}$). 
It is a barred spiral galaxy \citep{Sandage1987} 
and contains a number of superclusters \citep[e.g.][]{Zurita2001}
seen in {\it HST} images. 
NGC 7479 has been observed two times with {\it Chandra} and two times with 
{\it XMM-Newton}. It hosts several ULXs and we find that one of them 
is spatially coincident with a supercluster. The source is currently
not found in the {\it Chandra Source Catalog}, and we label it
CXOU J230453.0+121959 according to the {\it Chandra} naming 
convention\footnote{http://cxc.harvard.edu/cdo/scipubs.html\#NAME} 
based on the position of the source.

\subsection{The position of the source.}
\begin{figure}
\resizebox{\hsize}{!}{\includegraphics[angle=0]{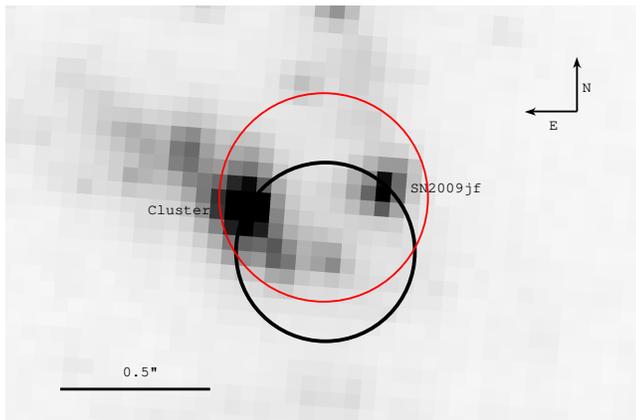}}
\caption{The position of ULX CXOU J230453.0+121959 on the optical
{\it HST} {\it F814W} image. The best X-ray position of ULX CXOU J230453.0+121959 from
{\it Chandra} OBSID 11230, indicated by the thick black circle. The position
in OBSID 10120, after the explosion of SN2009jf, is indicated by
the thin red circle. The positions were determined 
by direct matching of the X-ray and {\it HST} images using one matching 
source.}
\label{fig:position}
\end{figure}

The absolute astrometry of both the {\it Chandra} and the {\it HST} images
are uncertain at the $\sim1$ arcsec level. We have therefore
searched for sources that are both observed in X-rays and in
the optical, and have a reliable position in both. One such source 
exists, a bright foreground star.
While this is clearly not an ideal case, the distance of the
reference source to ULX CXOU J230453.0+121959 is relatively small 
($\sim23$ arcsec), and it is therefore possible to use it to precisely 
match the X-ray images to the {\it HST} images. We use the
{\it HST ACS WFC F814W} image jb4u53020 for the matching.
As both the {\it HST} and the {\it Chandra} images are well-calibrated,
we assume that there is no distortions between the two images, and
the rotation of the {\it HST} image was corrected using {\it 2MASS}
counterparts. The positions of the {\it} Chandra images were then
shifted to match the reference source in the {\it HST} image.
In figure \ref{fig:position} we show the position of the X-ray source
on the {\it HST} image. The thick black circle indicate the 
position found from {\it Chandra} OBSID 11230 (having the best statistics), and
and the thin red circle indicates the position found from {\it Chandra}
OBSID 10120.
The size of the circles indicate the statistical error on the
position of the X-ray source and the boresight correction
($\sim 0.3$ arcsec). The positions of the stellar cluster and the
ULX are consistent within the 1 sigma confidence limits.
The ULX is therefore most likely related to the cluster, even if it
could likely be located outside of it (similar to the other three
ULXs discussed above).
An older set of {\it HST WFPC2} observations cover the same region.
Comparison with these shows that the foreground reference star has
a small proper motion $\sim0.01$ arcsec per year. The {\it HST ACS}
images used for the matching are taken $\sim$1 year after the X-ray
observations, and the error due to this is therefore negligible.
 
We have verified the position using other optical intermediate 
images/catalogues, covering larger areas. These have therefore
more matching sources, but are less precise. The maximum number
of matching sources found was 4 comparing the 11230 X-ray source
list with 2MASS. This yields a position 0.2 arcsec directly
north of the position found from the direct {\it HST} matching
using OBSID 10120 (red circle in figure \ref{fig:position}), with
an error of $\sim 0.5$ arcsec. 

\subsection{The properties of the supercluster}
Below we summarize the important parameters of the cluster.
The metallicity of the surrounding region is found
to be between solar and LMC \citep{Valenti2011}, and for estimating
the physical parameters we assume a metallicity of Z=0.012.
The super-cluster is observed in two archival {\it HST} {\it WFPC2} images
taken with the {\it F569W} and the {\it F814W} filters (similar to the {\it V} 
and
{\it I} band filters, respectively).
The analysis of \citet{Valenti2011} gives
VEGA magnitudes, corrected for Galactic extinction, of $m_{F569}=19.84$
and $m_{F814}=19.46$ (absolute magnitudes $M_{F569}=-12.81$ and 
$M_{F814}=-13.19$). 
Unfortunately the two bands are not very
useful for inferring the age of the stellar association. A
comparison of the colours with simple stellar population
models \citep{Girardi2000,Marigo2008} gives an age between 10 and
100 Myr. The total mass of the stellar population depends on
the age and is between $7\times10^{5}$M$_{\odot}$ and $1.8\times10^{6}$
M$_{\odot}$. From the {\it HST ACS WFC F814} image jb4u53020, we find
that half of the mass is contained in a single unresolved central
cluster. The rest of the mass belongs to a collection of smaller
clusters scattered along the SW-NE axis. We note that while the surrounding
conglomerate of clusters is also young, but not necessarily co-eval with
the super-cluster.
One pixel corresponds to 8 pc and the {\it PSF} FWHM is $\sim20$ pc 
at the distance of NGC 7479. This
is smaller than the typical distance between individual clusters
in OB associations \citep[e.g.][]{Reipurth2008a,Reipurth2008b}, and the
central cluster is therefore unlikely to be a conglomerate of smaller 
clusters.
However, with the current spatial resolution, it is not possible to say 
whether the central cluster is compact enough to be expected to
form an IMBH \citep{Portegies2004}.

\subsection{The X-ray properties of the source}
\begin{figure}
\resizebox{\hsize}{!}{\includegraphics[angle=0]{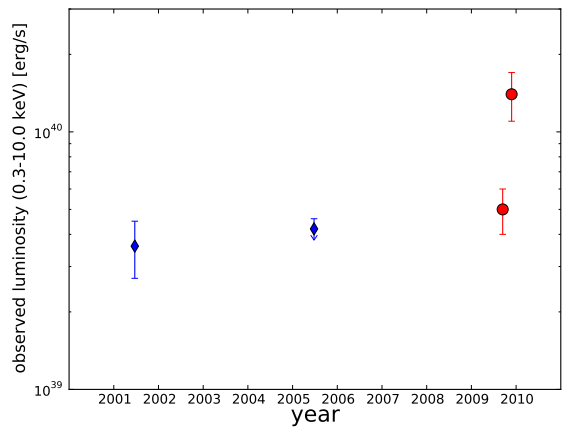}}
\caption{The X-ray lightcurve of CXOU J230453.0+121959. The diamonds
indicate {\it XMM}-observations and the circles {\it Chandra} observations. 
The arrow indicates that the second point is an upper limit.}
\label{fig:lightcurve}
\end{figure}
In the two {\it Chandra} observations, the source is bright enough to
extract the spectra, but the number of counts in the best observation
is only $\sim$120. We therefore use the Cash-statistics for spectral
fitting with minimum 5 counts in each bin (needed for the Cash
statistics to provide a reliable goodness-of-fit estimate). For
both observations a simple absorbed power-law provides a good fit to
the spectra, but both the slope and the absorbing column is poorly
constrained. We therefore fix the absorption to the Galactic
foreground value of 5.1$\times10^{20}$ cm$^{-2}$ \citep{Dickey1990}.
There is evidence for a varying spectrum, with a
best-fit power-law slope of $1.73\pm0.19$ in OBSID 11230 
and $1.13\pm0.19$ in OBSID 10120.
However, the significance of this is marginal $\gtrsim2\sigma$. 
As the spectral
models are poorly constrained, it is not possible to infer the
unabsorbed luminosities, and we instead characterize the sources
by the absorbed luminosities, noting that absorption is often
important in ULXs, so the unabsorbed luminosities might be several
times higher.
The luminosity of the source is clearly different in
the two observations, being $1.4\pm0.29\times10^{40}$ in the later 
observation and $5.0\pm1.0\times10^{39}$ erg s$^{-1}$ in
the earlier, but well above the Eddington limit of a typical
stellar mass black hole in both observations.

Two {\it XMM-Newton} observations from 2001 (OBSID 0025541001) and 2005
(OBSID 0301651201) cover the same region. 
The source is not detected through the standard data processing 
(PPS source lists), but a low-significance source is apparent in 
the 2001 observation. No source is visible by eye in the 2005 observation.
There is a nearby brighter source, and contamination is therefore
a serious issue. We therefore perform photometry using an
encircled energy fraction of 50 per cent. The source is detected with
a (corrected) EPIC PN (medium filter) count rate of 0.010 cts s$^{-1}$ 
in observation 0025541001, 
corresponding to an absorbed luminosity of $\sim3.6\pm0.9\times10^{39}$ 
erg s$^{-1}$,
assuming the spectral shape found in {\it Chandra} OBSID 11230.
There is no source in observation 0301651201. There we determine the 3$\sigma$
upper limit on the EPIC PN (thin filter) count rate to be 0.012 cts s$^{-1}$, 
corresponding to an absorbed luminosity of $4.2\times10^{39}$ erg s$^{-1}$.

ROSAT covered the region in the all-sky survey, but not with
any pointed observations. CXOU J230453.0+121959 was not detected in
the survey, but neither were nearby brighter ULXs.

\section{Discussion}
Out of the several hundred currently known ULXs \citep{Liu2005,Walton2011},
only four (including CXOU J230453.0+121959) have been associated with
young massive stellar clusters. The two Antennae ULXs have been
shown to be displaced from the stellar clusters \citep{Clark2011},
and above we have shown that this is also the case for M82 X-1.
We have found a new ULX, CXOU J230453.0+121959, that is associated 
with a supercluster. However, the displacement found for the other
three ULXs is also possible for this new ULX. Below we first
discuss the possible relation between CXOU J230453.0+121959 and
SN2009jf, and then we discuss the implications of the low number
of ULXs found in clusters and their displacement.

\subsection{Is the ULX related to supernova SN2009jf}
The latest X-ray observation, {\it Chandra} OBSID 10120 is taken shortly
after the explosion of the type Ib supernova SN2009jf at a distance
of $\sim0.3$ arcsec from the position found from {\it Chandra} OBSID 11230.
The luminosity of the ULX is much higher in this observation, compared to the
three other X-ray observations.
In some cases type Ib/c supernovae have been shown to be very bright in X-rays 
\citep[up to $\sim10^{40}$ erg s$^{-1}$][]{Sutaria2003,Chevalier2006,Immler2008}
within the first months after the supernova explosion, comparable to 
(but never as bright as) the
observed luminosity of ULX CXOU J230453.0+121959 in {\it Chandra} OBSID 10120.
This could indicate that the brightness is actually caused by the
supernova explosion. This is also consistent with the hardening 
of the spectrum, which is opposite to the behavious seen in most ULXs
\citep[although some ULXs do show such transitions][]{Fabbiano2003}.
However, the luminosity changes seen in ULX CXOU J230453.0+121959 
are common in other ULXs \citep{Fabbiano2003}, and the spectrum is 
still softer than what is typically seen shortly after the supernova explosion
\citep{Chevalier2006}.

A chance superposition is not unlikely, given that the large
area covered by the error region of an X-ray source at this distance, 
and the fact that CXOU J230453.0+121959 and SN2009jf might have been 
born in the
same cluster (or at least in the same association of clusters). 
It is even possible that CXOU J230453.0+121959 could be related to
SN2009jf. Indeed it is believed that some type Ib supernovae come
from binaries \citep{Podsiadlowski1992}, 
where a massive progenitor star transfers mass to a neutron
star or a black hole, causing the emission of X-rays. Indeed such
a candidate progenitor was proposed for the type Ib supernova
SN2010O \citep{Nelemans2010}. The connection between ULXs and
star-forming regions \citep{Irwin2004} indicates that the donor
stars in ULXs are high-mass stars transferring mass to a BH,
similar to the class of supernova type Ib with HMXB progenitors.

We note that both if CXOU J230453.0+121959 is related
to SN2009jf and if not, the short distance to the 
supercluster is interesting. The discussion below is valid in both
cases.

\subsection{Single star remnant}
If the ULX is the remnant of a single star, the radiation would
either have to be beamed, or the BH mass would have to be very high.
At low metallicities ($\lesssim0.4$Z$_{\odot}$), 
mass loss from the massive star progenitors of
black holes is expected to be inefficient
\citep{Maeder1992,Heger2003,Vink2005}. This can lead to very high
core masses at the end of their life, and they can therefore form
black holes with masses up to $\sim100$M$_{\odot}$. It is therefore
possible to form ULXs through normal binary evolution, without invoking
beamed radiation. However, the environment of CXOU J230453 is found 
to have a metallicity only slightly below solar, and if the current 
understanding of stellar evolution is correct, it is not possible
to form black holes more massive that $\sim15$M$_{\odot}$ in this
environment \citep{Heger2003}.

The continued detection of the source at luminosities above few times
$10^{39}$ erg s$^{-1}$ over several years shows that it is not a short
super-Eddington outburst. The highest reliable (ignoring the possibly
contaminated observation) observed absorbed X-ray luminosity is $5\cdot10^{39}$
erg s$^{-1}$. It is difficult to extrapolate this, in particular since there
are no real constraints on the absorbing column, but the bolometric
luminosity is likely $\gtrsim2$ times the X-ray luminosity in the {\it Chandra}
energy band. Sources emitting at
super-Eddington luminosities are known in the Milky Way. GRS1915+105
has been found to occasionally reach a luminosity in excess of $7\times10^{39}$ erg s$^{-1}$ \citep{Greiner1998}, similar to CXOU J230453.0+121959, and
the black hole accretor is 14M$_{\odot}$ \citep{Greiner2001}. 
However, the duration of this very bright state is only $\sim10$ s, and
the average luminosity over a few hours around this state is a factor
of $\sim3$ lower. From the {\it RXTE ASM} lightcurve we find that 
GRS1915+105 spends less than 1 per cent of the time
at luminosities above $10^{39}$ erg s$^{-1}$, and is therefore not
a good comparison for CXOU J230453.0+121959 and similar ULXs.

As discussed above the association of ULXs with massive clusters is
rare. Also HMXBs with sub-Eddington luminosities are rarely found in
clusters \citep[e.g.][]{Kaaret2004}. There are two reasons for this. 
Less dense clusters can
dissolve before the formation of HMXBs, 
and most HMXBs are ejected from their host clusters, either due
to the high velocities imparted by the supernova explosion, or
due to dynamical interactions in the cluster.

\subsection{IMBH}
In this scenario stellar interactions in a dense cluster causes
the massive stars to fall to the center on a timescale that is
shorter than their evolutionary timescale. Once accumulated there,
they collide and form a massive ``runaway star'', which most likely
collapses directly into a black hole without significant mass loss
\citep{Ebisuzaki2001,Portegies2002}. It should be noted that the evolution of such 
massive stellar objects is highly uncertain, and it is therefore not 
clear that an IMBH is actually formed in this way.
To become a ULX the black hole must capture a companion star, which
is natural in the dense environment in the centre of the cluster.
In this scenario it is therefore natural to find a ULX associated
with a massive cluster. However, several hundred ULXs are known
\citep[e.g.][]{Liu2005} only four are possibly related to massive
young clusters, and of these CXOU J230453.0+121959 is the only one
that could still be inside the cluster. In the IMBH
scenario, it is hard to explain why so few ULXs are associated with 
clusters, as they are dense enough to remain bound (and evolve into 
globular clusters). It is possible that most of the IMBHs are ejected 
from the clusters. The short distances ($<$few tens of pc) to the clusters 
makes this a likely explanation for M82 X-1, and the two Antennae
ULXs, and also CXOU J230453.0+121959 if it is found to be outside the
cluster. Such distances can be reached within a Myr, even with moderate
velocities of tens of km s$^{-1}$. Once ejected
the IMBHs are unlikely to capture companion stars, and they
must therefore accrete from companions that were ejected together
with the IMBHs from the clusters. On the other hand, the observations
of ULXs in globular clusters \citep{Maccarone2007,Maccarone2011} indicates that
some IMBHs might be retained in these clusters. If these indeed are
IMBHs, it is puzzling why they are not seen in the early phases of
the clusters.

\section{Conclusions}
We have re-analyzed the position of M82 X-1 and
found that the X-ray source is offset from the cluster with a confidence
of 3 sigma.
We have also identified a new ULX, CXOU J230453.0+121959, that is
associated with a young massive supercluster of a mass 
$\sim10^{6}$M$_{\odot}$. The X-ray source is observed three times
over 9 years with {\it Chandra} and {\it XMM-Newton}, but is probably
contaminated by SN2009jf in the last observation.
CXOU J230453.0+121959 is therefore the currently best
candidate for a ULX inside a massive young cluster.  
The rarity of observing ULXs inside massive clusters makes
it unlikely that most ULXs are formed inside clusters, unless 
they are kicked out of the clusters at birth.

\section*{Acknowledgments}
This research is supported by NWO Vidi grant 016.093.305.
This research has made use of data obtained from the {\it Chandra} 
Data Archive, 
and software provided by the {\it Chandra} X-ray Center (CXC) in the application 
package CIAO. This research is based on observations obtained with 
{\it XMM-Newton}, an ESA science mission with instruments and contributions 
directly funded by ESA Member States and NASA. Based on observations 
made with the NASA/ESA {\it Hubble Space Telescope}, obtained from the data 
archive at the Space Telescope Institute. STScI is operated by the 
association of Universities for Research in Astronomy, Inc. under the 
NASA contract  NAS 5-26555.

\bibliographystyle{mn2e}
\bibliography{/home/voss/work/bibliography/general}

\label{lastpage}

\end{document}